\documentclass[aps,graphicx,superscriptaddress,12pt]{revtex4-1}

\usepackage{graphicx}

\newcommand{\CuC}{(Cu,C)Ba$_{2}$Ca$_{2}$Cu$_3$O$_{9+\delta}$}
\newcommand{\cuc}{(Cu,C)Ba$_{2}$Ca$_{3}$Cu$_4$O$_{11+\delta}$}
\newcommand{\CuCTarget}{Ba$_{2}$Ca$_{3}$Cu$_{4.6}$O$_{y}$}
\newcommand{\STO}{SrTiO$_{3}$}
\newcommand{\YBCO}{YBa$_2$Cu$_3$O$_{7-\delta}$}
\draft 

\begin{document}

\title{Superconducting thin films of (Cu,C)Ba$_{2}$Ca$_{2}$Cu$_3$O$_{9+\delta}$ with zero resistance transition temperature close to 100 K} 

\author{Meng-Jun Ou}
\author{Yuecong Liu}
\author{Yi Wang}
\author{Hai-Hu Wen}
\email{hhwen@nju.edu.cn}
\affiliation{National Laboratory of Solid State Microstructures
	and Department of Physics, Collaborative Innovation Center of Advanced Microstructures, Nanjing University, Nanjing 210093, People’s Republic of China}

\begin{abstract}
High superconducting transition temperature is favorable for the applications of superconductors. Some cuprate superconductors have the transition temperatures above 100 K, such as the Hg- or Tl-based 1223 and 1234 phases, but many of them contain the toxic elements, like Hg and Tl. Meanwhile, the high anisotropy makes the vortices easy to move, thus the irreversibility magnetic field is very low in the liquid nitrogen temperature region. Here we report the successful synthesis of superconducting thin films (Cu,C)Ba$_{2}$Ca$_{2}$Cu$_3$O$_{9+\delta}$ with the zero-resistance transition temperature reaching 99.7 K. The film shows a very good crystallinity with (00l) orientation. The superconducting transitions are rather sharp as revealed by both resistivity and magnetization measurements. Temperature dependent resistivity has been measured under different magnetic fields along c-axis and ab-plane, and the irreversibility lines have been achieved. The resistivity was also measured at different temperatures with the magnetic field rotated in the ac-plane, and the data can be nicely scaled by using the anisotropic Ginzburg-Landau model, yielding a temperature  dependent anisotropy which varies from 17 at 110 K to 4 at 77 K. We also measured the magnetization-hysteresis-loops and calculated the critical current density through the Bean critical state model. The critical current density at 77 K is about 6$\times$10$^5$ A/cm$^2$ (zero field), thus the film may be a good candidate for the applications of superconducting cables or the high frequency superconducting filters.
\end{abstract}

\pacs{}

\maketitle 

\section{Introduction}
Since the discovery of the cuprate high-temperature superconductors \cite{BednorzPB1986}, enormous efforts have been dedicated to explore practical superconductors with superior properties for applications. There are three important parameters for practical application of superconductivity: high critical temperature $T_c$, high critical current $J_c$, and high irreversibility field $H_{irr}$. The irreversibility line $H_{irr}$(T) marks the phase boundary separating the region for zero and finite resistivity, below this line zero resistivity state can be achieved. Superconducting materials with critical temperatures higher than the liquid nitrogen (LN$_2$) temperature are more attractive for industrial applications due to their low cost. Hundreds of cuprate superconducting materials have been discovered, including Hg-based, Tl-based, Bi-based and RE-based (RE is a rare-earth element, e.g. Y, Gd, Nd etc.) systems with superconducting transition temperatures higher than the LN$_2$ temperature\cite{SchillingN1993,ShengN1988,MaedaJJoAP1988, CavaNature1987}. Hg-based and Tl-based systems have transition temperatures beyond 100 K, but the presence of the toxic elements Hg and Tl greatly limits their applications. The superconducting transition temperature of the Bi-2223 phase is about 110 K\cite{FujiiPRB2002}. However, due to the strong anisotropy and very layered structure,  the irreversibility field in the LN$_2$ region is very low. Superconducting tapes based on the Bi-2223 material have been mainly used in superconducting cable application in LN$_2$ temperature region, or applications under magnetic fields at low temperatures. The \YBCO (YBCO) phase with the transition temperature close to 90 K is considered as the most promising superconducting material because of its relatively high irreversibility field at LN$_2$ temperature. However, the short superconducting coherence length causes the weak links between grain boundaries, which great limits the pass through of the superconducting current. Consequently, for YBCO superconductors, thin films must be fabricated by epitaxial technology to meet the industrial performance requirements. Current research is focused on enhancing the critical current density and reducing its cost.

Another nontoxic cuprate superconducting system with $T_c$ beyond 100 K is \\(Cu,C)Ba$_{2}$Ca$_{n-1}$Cu$_{n}$O$_{2n+3}$, which is isostructure with the Hg-based system. The synthesis was made by several research groups in the middle of 1990s \cite{KawashimaPCS1994,KawashimaPCS1994b,KumakuraPCS1994}. In 2018,  an extremely high irreversibility field was discovered in the \cuc((Cu,C)-1234) polycrystal \cite{ZhangSA2018}. Its irreversibility field is 15 T at 86 K and still 5 T at 98 K. This superior property makes the material much attractive. However, the (Cu,C)-1234 bulk can only be synthesized by high-pressure method and the size of the samples is only several millimeters, which greatly limits its large-scale applications. Several groups attempted to grow the (Cu,C)-1234 films by pulsed laser deposition (PLD) \cite{AllenPCS1995,CalestaniPCS1999,PrellierPCS1997} and molecular beam epitaxy (MBE) \cite{ShibataPCSaiA2006,ShibataJoCG2007}. But the superconducting critical temperature is much lower than that of bulk samples. Until 2020, (Cu,C)-1234 thin film with a transition temperature of 96 K has been synthesized by our group ~\cite{DuanSSaT2020, DuanPCSaiA2020}. Nevertheless, its critical current is lower compared with that of the bulk material due to the film quality, and there is enormous space to further increase the critical current density by optimizing the preparation conditions. For the \CuC ((Cu,C)-1223) bulk, the superconducting transition temperature of pristine samples was 67 K, and after annealing in argon gas it rised up to 120 K\cite{ChailloutPCS1996}. The irreversibility field varies under the different annealing conditions, and the sample with $T_c =111$ K has an irreversibility field of 9.4 T at 77 K\cite{IyoPB2000}, which is close to the best value of YBCO. However, just like (Cu,C)-1234, bulk materials of (Cu,C)-1223 can only be fabricated by high-pressure method. The superconducting transition temperature of the (Cu,C)-1223 film synthesized by MBE was only 42 K\cite{ShibataJoCG2007}. In our knowledge, there have been no reports about the thin film growth for (Cu,C)-1223 by other epitaxial methods. Thus it is highly desired to grow (Cu,C)-1223 films with higher transition temperature and explore possible applications.

In this work, we report the successful growth of the $c$-axis oriented (Cu,C)-1223 thin films by pulsed laser deposition (PLD) technique. We achieved films with zero resistance transition temperature ($T_{c0}$) of about 99.7 K. The temperature dependent resistance under different magnetic fields was measured. Using 1\% $\rho_n$ as a criterion, its irreversibility field at 77 K was calculated to be 4.6 T by interpolation when the magnetic field is parallel to $c$ axis. In order to calculate the anisotropy, we measured the angle-dependent resistance of the film at different magnetic fields. Based on the anisotropic Ginzburg-Landau (GL) theory, the calculated anisotropy is 4 at 77 K and 17 at 110 K. The critical current of the film calculated from the magnetization hysteresis loops (MHLs) reaches 6$\times$10$^5$ A/cm$^2$ at 77 K by the Bean critical state model. We found that~(Cu,C)-1223 has a low anisotropy and a high critical current. The results suggest that the material has a great potential for applications in the LN$_2$ temperature region.

\section{Material and methods}

\subsection{Thin film growth and characterization}
The \CuC\ thin films were grown on single crystal \STO (001) substrates by pulsed laser deposition technique using a KrF excimer laser ($\lambda$ = 248 nm). The target with the nominal composition \CuCTarget\ was prepared by solid-state reaction method. Stoichiometric amounts of BaCO$_3$, CaCO$_3$ and CuO powder were mixed, ground, pelletized, and then sintered several times at 800-860~$\rm{^\circ C}$ for 12-24 h. During the deposition, mixed gas with different partial pressures of O$_2$ and CO$_2$ was introduced into the chamber, and the pressure in the chamber was maintained at 15-25 Pa. The deposition temperature was 620-680 $\rm{^\circ C}$. The laser energy density and repetition rate were 1-2 J/cm$^2$ and 5~Hz, respectively. After the deposition, the film was cooled down to room temperature at a rate of $5 \rm{^\circ C}/min$ in the same atmosphere as the deposition.  The XRD measurements were done using a Bruker D8 Advance diffractometer. 

\subsection{Magnetization and resistivity measurements}
Electrical resistance with magnetic fields from 0 to 11 T was measured using the standard four probe method in a physical property measurement system (PPMS-14T, Quantum Design). The angle dependent resistivity measurements were performed with the angle $\theta$ varied from $0 \rm{^\circ}$ to $180 \rm{^\circ}$,  where $\theta=0 \rm{^\circ}$  represents the magnetic field $H\parallel c$. The current was always perpendicular to the magnetic field and applied in the $ab$-plane. The magnetization measurements were carried out with a SQUID VSM-7T (Quantum Design).
\begin{figure*}
	\centering
	\includegraphics[width=\textwidth]{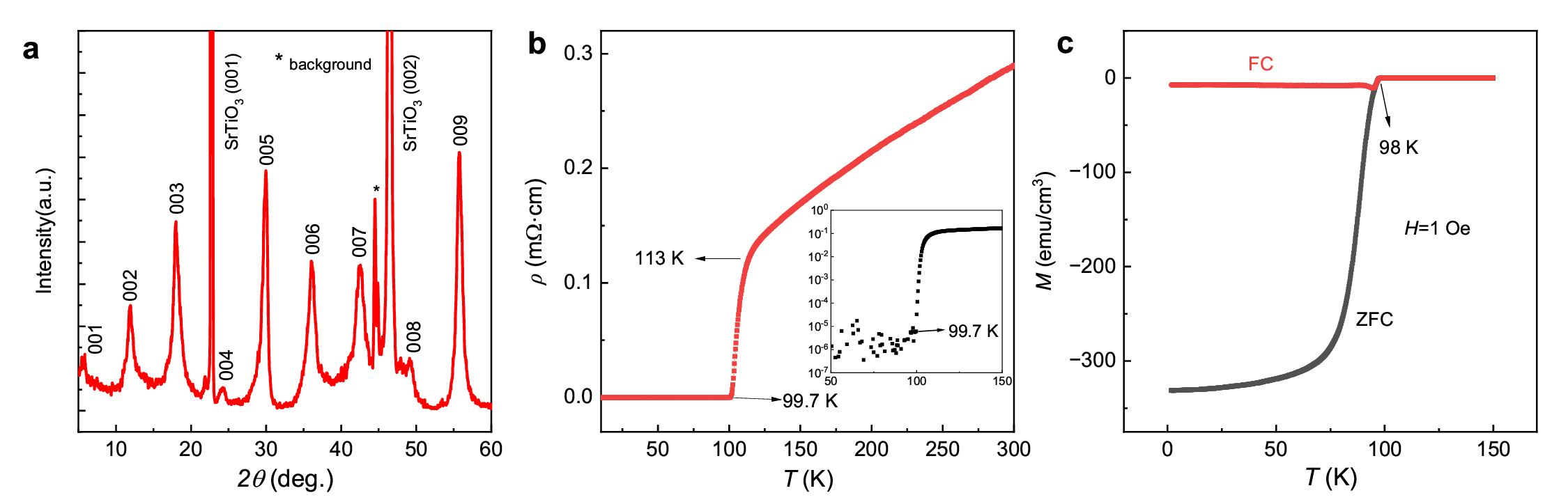}
	\caption{ X-ray diffraction (XRD) pattern, magnetic and transport properties of the \CuC\ film. a) X-ray diffraction pattern of the \CuC film on \STO (001) substrate. b) Temperature dependent resistivity curve measured at zero magnetic field. The inset shows the data near $T_c$ in the semilogarithmic scale. c) The temperature dependence of magnetic susceptibility under a magnetic field of 1 Oe.}
	\label{fig1}
\end{figure*}
\section{Results and discussion }

Fig.~\ref{fig1}a shows x-ray diffraction (XRD) pattern of the \CuC\ film. From the XRD pattern, one can see only the (00l) diffraction peaks without any obvious impurity peak. Analysis on the XRD data indicates that the film is $c$-axis orientation with $c$=14.83 \r A. The $c$-axis lattice constant is slightly larger compared with the as-prepared bulk material of the \CuC\ phase, probably due to the difference in oxygen or carbon content. We must emphasize that this is the first report about the successful synthesis of $c$-axis oriented (Cu,C)-1223 films.  In Fig.~\ref{fig1}b, the resistivity curve shows that the zero resistance transition temperature of the film is 99.7 K (indistinguishable from the noise background as shown in the inset). And the onset transition temperature is about 113 K. The resistivity exhibits metallic behavior in the normal state, which is similar to that reported for the bulk samples. However, in an earlier literature ~\cite{ShibataJoCG2007}, the synthesis of (Cu,C)-1223 thin films with the $a$-axis orientation grown by MBE was reported, the zero resistance transition temperature was $T_{c0}$  = 42 K, and a semiconductor behavior was observed in the normal state. Our work here demonstrates a significant improvement in the quality of the film. It should be noted that the as-prepared  (Cu,C)-1223 bulk samples are in the overdoped state as reported by several groups~\cite{ChailloutPCS1996, IyoPB2000,KodamaPCS2003}. And the $c$-axis lattice constant increases from 14.766 \r A to 14.825 \r A while the superconducting transition temperature is raised from 67 K to 120 K by a post-annealing with reduction of oxygen. This phenomenon can also be found in \YBCO \cite{CavaNature1987}. Thus it is reasonable that the (Cu,C)-1223 film grown at low partial oxygen pressure exhibits higher critical temperature and larger $c$-axis lattice constants. Fig.~\ref{fig1}c presents the temperature dependence of magnetic susceptibility measured in zero-field-cooled (ZFC) and field-cooled (FC) modes with the magnetic field (1 Oe) perpendicular to the film surface. The M-T curve shows that the superconducting transition temperature is about 98 K, which is close to the zero resistance transition temperature measured from the resistivity. Due to the demagnetization effect, the shielding volume fraction calculated at 10 K is much greater than 100\%. These results show that the critical temperature of (Cu,C)-1223 film is significantly enhanced from 42 K to 99.7 K and there is still a space to improve that.
\begin{figure*}[ht]
	\includegraphics[width=\linewidth]{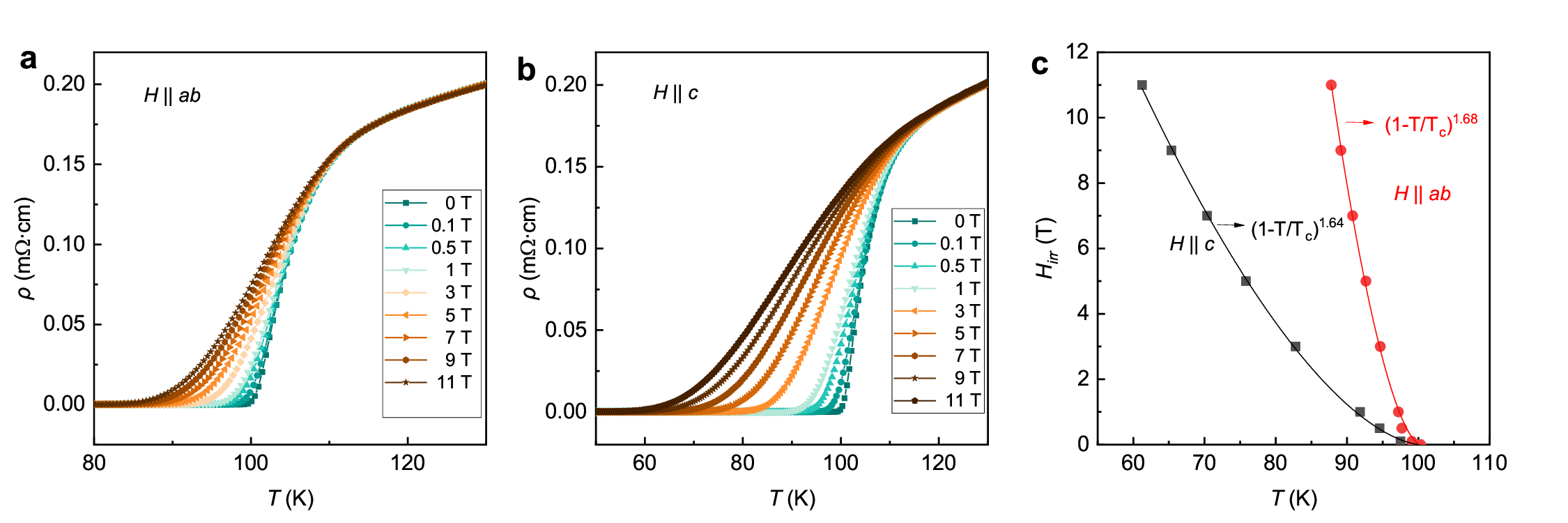}
	\caption{(a-b) Temperature dependent resistivity of the (Cu,C)-1223 film under external magnetic fields $H\parallel c$ and $H\parallel ab$, respectively. (c)  Temperature dependence of the irreversibility field $H_{irr}$. The solid lines show the fitting results by $H_{irr}=H_{irr}(0) (1-T/T_c)^{\beta}$ with $\beta=1.64$ for $H\parallel c$ and $\beta=1.68$ for $H\parallel ab$.}
	\label{fig2}
\end{figure*}

The temperature dependent resistivity of the \CuC\ film was measured under external magnetic fields $H\parallel c$ and $H\parallel ab$ as shown in Fig.~\ref{fig2}a-b. It can be seen that the superconducting transition gradually becomes broader as the magnetic field increases, and  the film exhibits a highly anisotropic response for different magnetic field orientations. When the external magnetic field is parallel to the $ab$-plane, the superconducting transition is slowly suppressed. And $T_{c0}$ is about 81 K at a magnetic field of 11 T. The irreversibility field can be calculated by using a criterion of 1\% $\rho_n$ , as shown in Fig.~\ref{fig2}c. The data can be well fitted by the power exponent $H_{irr}=H_{irr}(0) (1-T/T_c)^{\beta}$. For $H\parallel ab$, the exponent $\beta=1.68$. The irreversibility field is about 20 T at 77 K by extrapolation. For $H \parallel c$, the exponent $\beta=1.64$, which is very close to the YBCO film \cite{Haenisch2007}. And the irreversibility field is about 4.6 T at 77 K.  The irreversibility field of the (Cu,C)-1223 bulk sample with $T_c$ = 111 K is 9.4 T at 77 K and can be enhanced by controlling the doping level through annealing in N$_2$, suggesting that (Cu,C)-1223 may have a higher irreversibility field line than YBCO. 

\begin{figure*}[ht]
	\centering
	\includegraphics[width=0.7\textwidth]{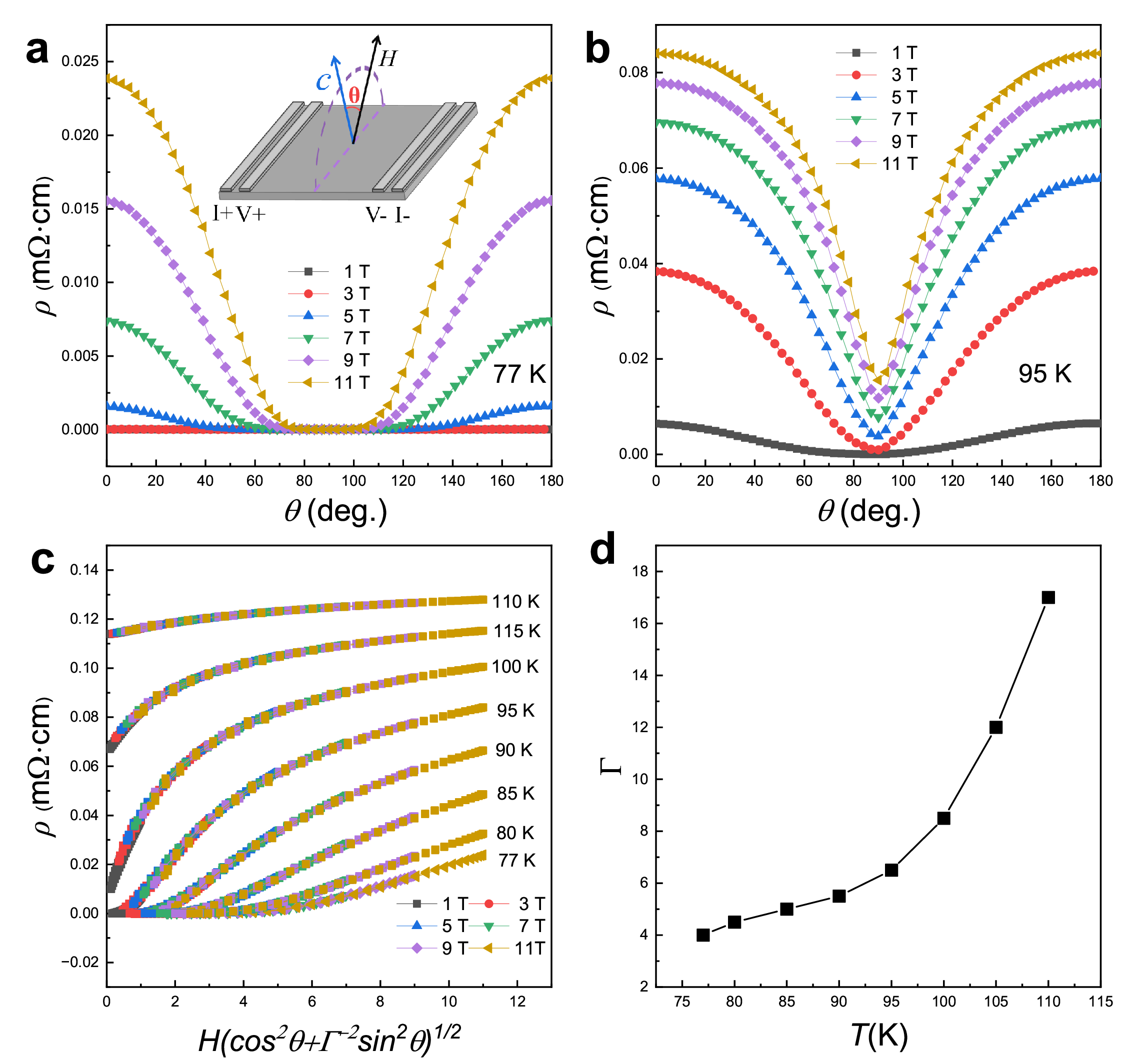}
	\caption{The angle dependent resistivity and anisotropy of the \CuC\ film. (a-b) Angle dependence of resistivity under different magnetic fields at 77K and 95K, respectively. The inset of (a) illustrates the definition of angle $\theta$. (c) The resistivity as a function of the effective magnetic field at different temperature. (d) The temperature dependence of anisotropy. }
	\label{fig3}
\end{figure*}

In order to have an assessment on the anisotropy of the system, we carried out the resistivity measurements under different magnetic fields with the field rotated in the $ac$-plane, allowing the applied magnetic field always perpendicular to the applied current direction. The superconducting anisotropy, defined as $\Gamma=\sqrt{m_c/m_{ab}}$ with $m_c$ and $m_{ab}$ the effective mass matrix element in the normal state, can be determined from the ratio of upper critical field $H_{c2}^{ab}/H_{c2}^c$. However, it might depend on the criterion adopted for defining the $H_{c2}$. In order to obtain a more reliable anisotropy of the (Cu,C)-1223 film, we measured the angle-resolved resistivity under different magnetic fields from 1 T to 11 T. According to the anisotropic Ginzburg-Landau theory, the angle-dependent upper critical field $H_{c2}^{GL}(\theta)$ is given by
\begin{equation}
	H_{c2}^{GL}(\theta)=\frac{H_{c2}^{c}}{\sqrt{cos^2(\theta)+\Gamma^{-2}sin^2\theta}}.
\end{equation}
Blatter et al~\cite{Blatter1992} proposed that the angle dependence of resistivity can be scaled as $\rho=\rho_0 f(H/H_{c2}^{GL})$. Therefore by adjusting the anisotropy $\Gamma$ and using the effective field $\tilde{H}=H\sqrt{cos^2\theta+\Gamma^{-2}sin^2\theta}$ as $x$-coordinate, the resistivity measured at different magnetic fields but a fixed temperature should collapse onto one curve. Fig.~\ref{fig3}a-b display the angle dependence of resistivity at $T$ = 77 K and 95 K. Fig.~\ref{fig3}c presents the scaling curves at different temperatures. One can see that all curves can be well scaled, which validates the model used here. In the scaling procedure, only one fitting parameter $\Gamma$ is involved, so the results are more reliable compared with the calculated values by the ratio of $H_{c2}^{ab}/H_{c2}^c$. Thus, the temperature dependence of the anisotropy $\Gamma$ can be obtained and the results are shown in Fig.~\ref{fig3}d. One can see that the anisotropy increases from 4 at 77 K to 17 at 110 K. And the anisotropy $\Gamma$ is 5.5 at 90 K, which is slightly larger than that in (Cu, C)-1234 single crystals \cite{HeMTP2022}. This is consistent with previous research that the superconducting anisotropy can be reduced by increasing the number of  the superconducting layers\cite{IharaPhysC1997}. 

\begin{figure*}[ht]
	\includegraphics[width=\linewidth]{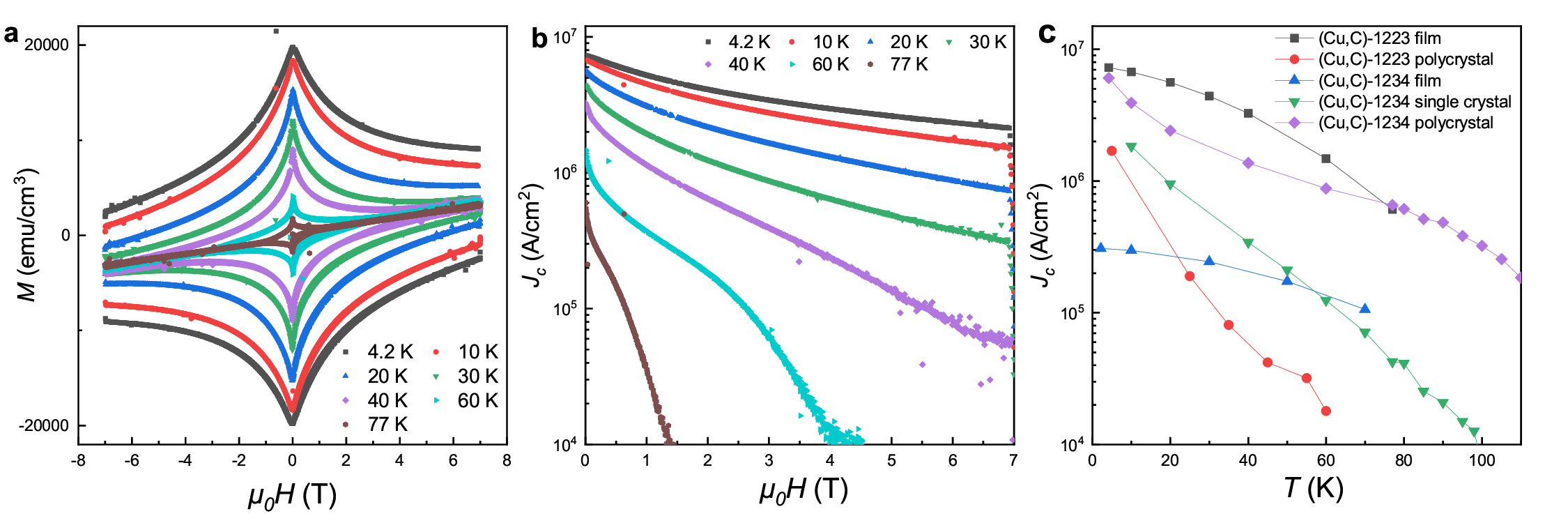}
	\caption{(a) The magnetization hysteresis loops measured at different temperatures with H$\parallel$ c. (b) The magnetic field dependence of the calculated critical currents by the Bean critical state model. (c) The critical current density $J_c$ at 0 T for (Cu,C)-1223 film, (Cu,C)-1223 polycrystal\cite{KumakuraPCS1994}, (Cu,C)-1234 film \cite{DuanSSaT2020}, (Cu,C)-1234 single crystal \cite{HeSSaT2021} and (Cu,C)-1234 polycrystal \cite{ZhangSA2018}.}
	\label{fig4}
\end{figure*}

Another important parameter for practical applications of superconducting materials is the critical current density $J_c$. As mentioned above, the (Cu, C)-1223 film has a relatively low anisotropy and may have a high irreversible line after further optimization. In order to evaluate the critical current density at different temperatures, the magnetization hysteresis loops were measured. Fig.~\ref{fig4}a shows the magnetization hysteresis loops of the film at different temperatures with $H\parallel c$. It is clear that the magnetization hysteresis loops are irreversible below a certain temperature and magnetic field. In addition, a paramagnetic background can be observed, which may originate from the substrate. A similar phenomenon was observed in (Cu,C)-1234 film\cite{DuanSSaT2020}. By using the Bean critical state model, the critical current density $J_c$ can be calculated from the magnetization hysteresis loops with the formula $J_c=20\Delta M/[w(1-w/3l)]$. Here $\Delta M$ with the unit of emu$/$cm$^3$ is the width of MHLs; $w$ and $l$ are the width and length of the sample. Fig.~\ref{fig4}b presents the  magnetic field dependence of the critical current density $J_c$.  It is clear that the calculated $J_c $ (0 T) reaches 7.2$\times$ 10$^6$ A/cm$^2$ at 4.2 K, which is greater than the value in (Cu,C)-1223 polycrystal. At 77 K, the critical current density $J_c$ reaches 6$\times$ 10$^5$ A/cm$^2$ at 0 T, which is one order of magnitude higher than that of (Cu,C)-1234 single crystals. Fig.~\ref{fig4}c shows the comparison of the temperature dependent critical current density at 0 T for (Cu,C)-1223 film, (Cu,C)-1223 polycrystal, (Cu,C)-1234 film, (Cu,C)-1234 single crystal and (Cu,C)-1234 polycrystal. It is evident that the (Cu,C)-1223 film has the highest critical current density at 4.2 K and 77 K among these materials. These results indicate that (Cu,C)-1223 films exhibit good current carrying ability. Furthermore, the introduction of artificial pinning centers into YBCO was identified as an effective method to enhance the critical current density and the irreversibility field \cite{Haugan2004, Varanasi2008, Matsumoto2009}. Thus it is possible to further improve the performance of  (Cu,C)-1223 by using the same technique.

\section{Conclusions}
The $c$-axis oriented \CuC\ thin films were for the first time successfully synthesized on \STO\ (001) substrates by pulsed laser deposition. The zero resistance transition temperature reaches 99.7 K, which is a significant improvement compared with previous reports. By using the anisotropic Ginzburg-Landau model, the temperature dependent anisotropy is determined from resistivity with angle-dependent magnetic fields. The anisotropy $\Gamma$  is about 4 at 77 K and about 17 at 110 K. In order to study the current carrying ability of the film, we measured the MHLs under different magnetic fields. The calculated critical current density J$_c$(0 T) is 7.2$\times$10$^6$ A/cm$^2$ at 4.2 K and 6$\times$10$^5$ A/cm$^2$ at 77 K. These results indicate that the \CuC\ material has great potential for applications in the liquid nitrogen temperature region.

\section*{Notes}
Chinese patent 202410414918.4

\section*{Acknowledgements}
We appreciate the kind help in the early establishment of the PLD  equipment by Hai-Feng Chu and Tianfeng Duan. The project is supported by the National Key Research and Development Program of China  (No. 2022YFA1403201), National Natural Science Foundation of China (Nos. 11927809, 12061131001).

\bibliographystyle{unsrt}
\bibliography{CuC1223}

\end{document}